\documentclass[twocolumn,amsmath,amssymb,showpacs,fleqn]{revtex4}
\usepackage{graphicx}
\begin{document}

\title{Light Scattering Spectroscopy:  A New Method for Precise Determination of Atomic Matrix Elements}
\author{M.D. Havey}

\affiliation{Old Dominion University, Department of Physics, Norfolk, Virginia 23529 \
}

\begin{abstract}
A new approach to precise determination of  atomic matrix elements is described whereby  measurement of spectral locations of zeros in the Rayleigh scattering cross-section allows frequency-domain extraction of matrix elements in terms of a fiducial quantity. Illustrations are made for scattering off the ground state in Li and Cs.\\
\end{abstract}


\maketitle

\section{Introduction}
An outstanding problem in experimental spectroscopy is precise determination of matrix elements describing interaction of electromagnetic radiation with matter \cite{Tannoudji}.  In atomic physics, such matrix elements are important to a wide range of fundamental problems, ranging from analysis of parity violation measurements in atoms \cite{Noecker} to determination of the astrophysical atomic abundances \cite{Mihalas}.   In studies on atomic or molecular systems, measurement observables which determine the magnitude of matrix elements have only recently attained a relative precision of about $\sim 10^{-3}$ in the best cases \cite{Goebel,Ekstrom,Oates,Rafec,Volz,Meyer}.   In contrast, other quantities such as energy level separations ($\sim 10^{-12}$) \cite{Shao} and atomic masses ($\sim 10^{-11}$) \cite{DiFillipo} may be measured to an extraordinary, and much greater, relative precision.    Improvements in measurement techniques, coupled with increasing sophistication of theoretical calculations, have produced a fruitful theory-experiment synergy leading to significant advances in the field.

In this paper a new method to determine  atomic and molecular transition dipole matrix elements is described.  Generally, this method relies on two-photon light scattering processes, such as  Rayleigh or Raman scattering \cite{Tannoudji,Marcuse}.  The amplitude for these nonresonant processes depends on the contributions of the spectral distribution of  intermediate states, of which often  only a subset is practically important.  Interference of the  contributions as a function of the spectral location of  so-called virtual levels for the process produces a set of zeros in the amplitude.  For a given selected initial and final level, the locations of the zeros, and the associated intermediate one-photon resonances, determine the weak-field response of the atomic system.  With the method, the transition matrix elements connecting the various nondegenerate states of the system may, in principle, be determined to the same level of precision as a fiducial value.  The fiducial value may be derived, for example,  from the measured strength of a principal resonance transition.   By experimental location of related zeros in the Raman scattering or in the two-photon absorption spectrum, the relative phases of the matrix elements may also be found.  The method thus achieves a mapping of matrix elements into the frequency domain, where their values may be precisely determined.  It is important to note that these considerations apply generally to atomic, molecular, or solid state systems having a discrete energy spectrum.

Other authors \cite{Quattropani} have noted in special cases the existence of zeros in the two-photon excitation spectrum.  For example, Quattropani, $\emph{et al.}$ \cite{Quattropani} have calculated in atomic H the location of zeros for several s-p-s two-photon, two-color transitions, and argued that they are a general feature of that spectrum.  Experimental observation of zeros in the Rayleigh scattering spectrum due to fine- and hyperfine structure interferences have also been made in Na \cite{Tam,Walkup,Zei}.  Measurements of complete destructive interference in polarization-dependent, two-photon excitation have also been made on the 3s-3p-5s \cite{Meyer} and 3s-3p-4d \cite{Quattropani}  transitions in Na,  and on the 5s-5p-8s transition in Rb \cite{Beger}.   Rayleigh scattering cross-sections in the vicinity of the first resonance transition in Cs, including the location of the zero,  have been calculated by Penny \cite{Quattropani}.

In the remainder of this report, the method is presented in detail for the case of Rayleigh scattering  off an atomic ground level.  Results are presented for the zero locations, including the contributions of energetically remote  levels.  Numerical illustrations are made for the cases of  Li and Cs, for which precise fiducial values are available \cite{Ekstrom,Rafec}.  Finally, it is shown how the same approach may be used to find excited-state matrix elements, in terms of the same fiducial value.

\section{Analysis}
The fundamental expression for the Rayleigh scattering differential cross-section is well-known \cite{Tannoudji,Marcuse}, and is given, for scattering off the ground state,  by

\begin{equation}
\frac{d \sigma}{d \omega} = \frac{q^4 \omega^{\prime 3} \omega}{16 \pi^2 c^4 \hbar^2}|A|^2
\end{equation}

\noindent where the scattering amplitude is

\begin{eqnarray}
A = \sum_i \left(\frac{<f|e\cdot r|i><i|e'\cdot r|f>}{\omega_i + \omega'} \right. \\
\nonumber
\left. +  \frac{<f|e'\cdot r|i><i|e\cdot r|f>}{\omega_i - \omega}\right)
\end{eqnarray}

Here, $\omega$ and $\textbf{e}$ represent the frequency and polarization of the incident radiation, while $\omega^{\prime}$  and $\textbf{e}^{\prime}$  describe the scattered radiation.  The electric dipole operator is $q\textbf{r}$, where q is the electron charge and $\textbf{r}$ the electron position.  For elastic scattering, take $\omega = \omega'$ , thereby ignoring Doppler and recoil shifts.   Intermediate states are labeled as $\mid i>$, with energy  $\hbar \omega_i$, while the initial and final states are given as $\mid f>$.   The sum over intermediate states implicitly includes integration over the continuum.  This contribution is normally quite small, except for the case of H, and will be ignored here \cite{Bethe}.   In Eq. (2) the natural width $\gamma_i$  of the levels has also been ignored.  Thus the expression may not be used when the radiation frequency is very close to an atomic resonance.  However,  important cases occur when $\omega$ is far from atomic resonance, as measured in units of $\gamma_i$.

An essential feature of Eqs. 1 - 2 is that, as a result of destructive interference between contributions from the intermediate states, there exist frequencies $\omega_{oi}$ for which the scattering amplitude $A$ is zero. In addition, as the matrix element products in Eq. (2) are always positive when $\textbf{e}$ = $\textbf{e}^{\prime}$, there exists one zero located between each pair of excited-intermediate levels above the ground state.  When the natural widths are ignored, the scattering spectrum then consists of poles located at frequencies $\omega_i$  and a zero located between each pair $\omega_i$ and $\omega_{i+1}$.  These sets define the frequency response  of the atomic system initially in its ground level.  Finally, although the summation in Eq. (1) is over an infinite set of excited levels, the summation for Rayleigh scattering converges rapidly \cite{Weise}, so only a few levels contribute significantly.  Thus the sum may be truncated to provide a desired level of precision.  The roots of the resulting finite polynomial then constitute a set $\omega_{oi}$ which may be used to determine the matrix elements.

Note that to obtain the ultimate precision in this approach, the size of the truncated piece of the summation must be estimated.  This may be done by direct calculation of the bound level and continuum contributions as in recent calculations \cite{Theo} of the dipole polarizability of the ground state of atomic $Mg^+$ and $Ca^+$, for Which similar sums appear.  As in the present case, the summation giving the dipole polarizability is dominated by the resonance transition,. which contributes all but about 0.4 $\%$ for those cases.  As suggested by Beck and Nicolaides \cite{Beck}, an upper bound may also be placed on the truncated piece by combining existing measured or calculated values for the polarizability with precise measurements of the resonance line strength.

These observations may be used to provide a new method to precisely determine squared matrix elements, for transitions out of the ground level of many atoms.  To see this, note first that  the transition strength for excitation out of the ground level is normally concentrated in the lowest-energy-allowed transition \cite{Weise,Radzig}.  For example, in the alkali- and alkaline earth-metal atoms, the oscillator strength \cite{Definition1} of the first resonance transition is about f = 1, while $f_i$ for other transitions is typically 10 times smaller than this.  Thus the location of the zeros in the Rayleigh scattering  are spectrally close to the resonances (poles).    Second, recent experimental advances have permitted measurement of f for the first resonance transitions to a precision of about $10^{-3}$ in some cases.  This is to be compared to a typical precision of ~10 $\%$ for transitions to more energetic levels.  While other excited levels do contribute to the process, the contribution is small, and needs only be approximately known.  The spectral separation between an atomic resonance at $\omega_{i}$  and the associated zero at $\omega_{oi}$  is determined mainly by the matrix elements coupling the ground level to level i, relative to the matrix elements associated with the resonance transition.

\section{Illustrations}
\begin{figure}
\includegraphics[width=\columnwidth, keepaspectratio]{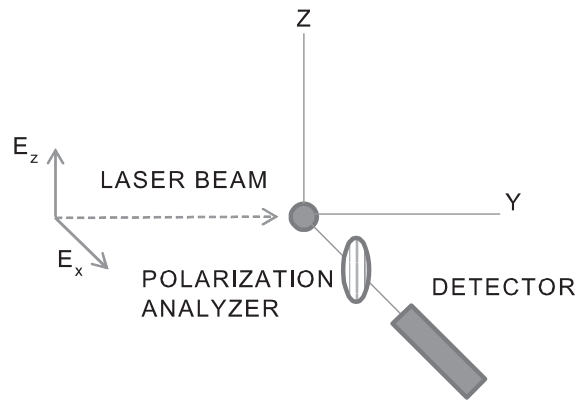}
\caption{ Schematic diagram of the geometry considered for polarization-dependent Rayleigh scattering.}
\label{Figure1}
\end{figure}

The  above considerations are  illustrated by means of a particular geometry and applied to Rayleigh scattering \cite{Definition2} off the $^2S_{1/2}$ ground level of Cs and Li.  Consider the situation shown  in Fig. 1, where the exciting radiation is taken to propagate along the y-axis, and is linearly polarized along either the z-axis or x-axis.  The scattered radiation is detected through a linear polarization analyzer oriented so as to transmit the component of the scattered radiation polarized along the z-axis.  Evaluation of Eq. (2) for frequencies in the neighborhood of a multiplet np gives

\begin{eqnarray}
A^n_{zz} = -\frac{1}{9} \left( \frac{2M_{n3/2}}{\Delta_n} + \frac{M_{n1/2}}{\Delta_n + \Delta^{fs}_n} \right.\\
\nonumber
\left. + \frac{2M_{n3/2}}{\Delta_n - 2\omega} + \frac{M_{n1/2}}{\Delta_n + \Delta^{fs}_n - 2\omega} - P^n_{zz} \right)
\end{eqnarray}

\begin{eqnarray}
A^n_{xz} = -\frac{1}{9} \left( \frac{M_{n3/2}}{\Delta_n} - \frac{M_{n1/2}}{\Delta_n + \Delta^{fs}_n} \right.\\
\nonumber
\left. - \frac{M_{n3/2}}{\Delta_n - 2\omega} + \frac{M_{n1/2}}{\Delta_n + \Delta^{fs}_n - 2\omega} - Q^n_{xz} \right)
\end{eqnarray}

In these expressions, $M_{nj}$ represents the squared radial transition matrix elements connecting the ground level with a particular level nj, where n is the principal quantum number and j the angular momentum quantum number.   They are defined as

\begin{equation}
M_{nj} = \left|\int_0^\infty R^*_{nlj}(r) R_{n^{\prime}l^{\prime}j^{\prime}}(r) r^3 dr\right|^2.
\end{equation}

where $l$ = 1, and $l$  = 0, are the orbital angular momenta of the excited and ground levels.  In the definition, the j-dependence due to angular momentum recoupling has been included in Eq. 3 - 4, leaving only intrinsic j-dependence due to relativistic mixing of the spin-orbit levels.   The detuning $\Delta_n = \omega  - \omega_{n3/2}$, is the offset of the radiation from resonance with the n $^2P_{3/2}$ level, while $\Delta^{fs}_n$ is the fine-structure splitting in the multiplet n.  The quantities $P_{nzz}$  and $Q_{nxz}$ represent the contributions, from all other levels, to the Rayleigh scattering amplitude:

\begin{eqnarray}
P^n_{zz} = \sum_{i\neq n} \left( \frac{2M_{i3/2}}{\omega_{i3/2} - \omega} + \frac{M_{i1/2}}{\omega_{i1/2} - \omega} \right.\\
\nonumber
\left. + \frac{2M_{i3/2}}{\omega_{i3/2} + \omega} + \frac{M_{i1/2}}{\omega_{i1/2} + \omega} \right)
\end{eqnarray}

\begin{eqnarray}
Q^n_{xz} = \sum_{i\neq n} \left( \frac{M_{ni3/2}}{\omega_{i3/2} - \omega} - \frac{M_{i1/2}}{\omega_{i1/2} - \omega} \right.\\
\nonumber
\left. - \frac{M_{i3/2}}{\omega_{i3/2} + \omega} + \frac{M_{i1/2}}{\omega_{i1/2} + \omega} \right)
\end{eqnarray}

As in Eq. 1, these amplitudes permit definition of a linear depolarization degree in terms of intensities $\sigma_{zz}$ and $\sigma_{xz}$:

\begin{equation}
P_L = \frac{\sigma_{zz} - \sigma_{xz}} {\sigma_{zz} + \sigma_{xz}}
\end{equation}

 \subsection{Application to Cs Rayleigh Scattering}
Using experimental data for Cs \cite{Radzig}, $P_L$ has been calculated as a function of $\Delta_7$  for Rayleigh scattering in the vicinity of the 6s $^2S_{1/2}$ $\rightarrow$  7p $^2P_{j}$  second resonance transitions.   Knowledge of matrix elements for these transitions is important for analysis of parity violation experiments in atomic Cs, and the present scheme is directly applicable to that case.   The results are presented in Fig. (2), and show the existence of two critical frequencies where $A_{zz}$ = 0 and $P_L$ = -100 $\%$.  For the data of \cite{Radzig}, these are located at about -43.25 $cm^{-1}$ and at -195.2 $cm^{-1}$, relative to resonant scattering from the 7p $^2P_{3/2}$ level.   Roughly speaking, the zero between the resonance transitions arises from interference between the scattering from the two 7p fine-structure components, while that to the low-energy side of the 6s $^2S_{1/2}$ $\rightarrow$  7p $^2P_{1/2}$  transition is due to the contribution of more energetically distant p-levels.   This contribution is dominated by the  6s $^2S_{1/2}$ $\rightarrow$  6p $^2P_j$ resonance transition, and so the location of the lower frequency zero is determined primarily by the very well-known ($< 10^{-3}$) 6p matrix elements, and the 7p matrix elements.  The much less-well-known 6s$^2S_{1/2}$ $\rightarrow$ np $^2P_j$  matrix elements (n > 7) contribute only weakly to the null location, and their contribution needs only to be estimated.  Thus, precise experimental determination of the $P_L$ = -100 $\%$ points directly allows extraction of the unknown quantities; the dipole matrix elements for the 6s$^2S_{1/2}$ $\rightarrow$ 7p $^2P_j$ transitions.  Previous polarization measurements \cite{Beger} in two-photon, two-color excitation in Rb have shown that precise determination of the spectral location ($ \sim \pm 10^{-2} cm^{-1}$) of the zeros is possible.  In addition, the Rayleigh scattering cross-section for these transitions is large, and the relative scattering intensity may be readily measured.   For this case, the matrix elements may be determined to $< 10^{-3}$, limited primarily by the precision of the fiducial resonance line oscillator strengths.

\begin{figure}
\includegraphics[width=\columnwidth, keepaspectratio]{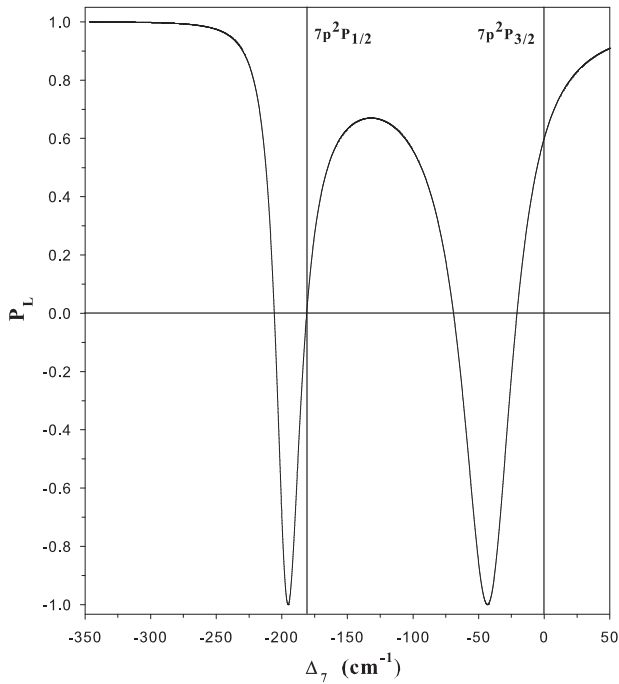}
\caption{Calculated linear polarization spectrum for Rayleigh scattering in the vicinity of the 6s $^2S_{1/2}$ $\rightarrow$ 7p $^2P_j$  (j = 1/2, 3/2)  second resonance transitions, illustrating the location  of the two zeros in $A_{zz}$ where $P_L$ = -1.0.}
\label{Figure2}
\end{figure}

\subsection{Application to Li Rayleigh Scattering}
\begin{table}[h!]
\caption{Tabulation of the location of zeros  in the atomic Li Rayleigh scattering cross-section for
 parallel linear polarization of excitation and scattered radiation.  The zeros are
 referenced to the associated 2s $^2S_{1/2}$ $\rightarrow$ np $^2P_j$  atomic resonance.  Fine-structure
 splitting is ignored.}
\begin{tabular}{c c c c}
  \hline
  n &  & $\omega_{oi}$ - $\omega_i$ ($cm^{-1}$)  &  \\
  \hline
    & Numerical & Numerical (2p only) & Approx. Analytical \\
  \hline
  3 & -38.37 & -38.07  & -38.44 \\
  4 & -33.10 & -32.96  & -33.17 \\
  5 & -25.24 & -25.43  & -25.29 \\
  6 & -16.70 & -17.15  & -16.72 \\
  \hline
\end{tabular}
\end{table}

The Li atom is also an interesting case, for it has the complexity of a three-electron system, and yet should be amenable to accurate theoretical  treatment.   This point is underscored by recent precise measurements of the lifetime of the 2s $^2S_{1/2}$ $\rightarrow$ 2p $^2P_j$ transition \cite{Ekstrom,Volz}, and the agreement of the measurements with sophisticated theoretical studies.  In this section results are given on the locations of  zeros, associated with the n = 3 - 6 levels,  in the Rayleigh scattering cross-section.

First, note that the spin-orbit coupling in Li is very small \cite{Radzig}.  Thus, except for the fine-structure splitting ($\sim 0.34 cm^{-1}$)  in the 2p $^2P_j$  multiplet, it may be ignored.  As the  amplitude $A_{nxz}$  is then nearly zero, the results are presented in terms of the locations of the zeros relative to the various np multiplet levels.  From Eq. 3 - 6, a useful and accurate expression for the zero location, relative to the np level, is

\begin{eqnarray}
\omega_{on} - \omega_n = \frac{2M_{n3/2}  + M_{n1/2}}{P^{n}_{zz}}
\end{eqnarray}
\begin{eqnarray}
P^{n}_{zz} = \sum_{i\neq n} \frac{ 2 \omega_i(2M_{i3/2}  + M_{n1/2})}{\omega_{i}^{2} - \omega^{2}}
\end{eqnarray}

In Table 1, results for $\omega_{on}$ - $\omega_n$  are presented as calculated by direct application of Eq. 3-7.  Also given are the approximate results from Eq. 9.  For illustrative purposes, the locations were obtained by using the atomic data from \cite{Radzig}.   First, note that the $\omega_{on}$ - $\omega_n$ are uncertain by more than 10 $\%$, limited by the current precision of the 2s - np oscillator strengths.  Second, by comparing $\omega_{on}$ - $\omega_n$ in Table 1 obtained with and without inclusion of levels other than 2p, it is seen that $P^n_{zz}$  is dominated by the contribution from the 2s $^2S_{1/2}$    $\rightarrow$ 2p $^2P_j$ resonance transitions, which are known to about 0.1 $\%$.  Thus, experimental determination of the zero locations determines directly the matrix elements in Eq. 9.  It is also evident from the table that contributions from more energetic p levels need only be known to 10 $\%$ or so, in order to obtain a precision  $\sim 10^{-3}$ in the squared matrix elements.   Of course, contributions of distant states may be more precisely determined by a series of measurements locating a set of zeros, followed by global fitting of the data.  Finally, we note that $\gamma_n$ /($\omega_{on} - \omega_n$) $<< 1$, so ignoring the natural width is justified in the cases considered.

\subsection{Application to determination of excited-level matrix elements}
Once transition matrix elements for the gs - np principal series are determined, it is possible to extend the scheme through a stepwise process in which a gs - np transition is $\emph{resonantly}$ excited with one light source.  Measurements of Rayleigh scattering off excited np levels are then made with a second light source.  There are  zeros in the np - n's and np - n"d scattering amplitudes, resulting in determination of  relative transition matrix elements as before.  The np - n's and np - n"d contributions may be distinguished through their  different dependence on light  polarization, even in the absence of spin-orbit interaction.  The relative measurements may be put on an absolute scale through the previously determined gs - np matrix elements, for the np - gs transition also contributes to the excited-level Rayleigh scattering.  This approach may evidently be extended to scattering off other excited levels, and matrix elements similarly determined.

\section{Conclusions}
A new method for determining transition matrix elements has been described.  The technique relies on measurement of Rayleigh scattering off  ground and excited levels of atoms.  Location of zeros in the scattering cross-section, or measurement of the polarization dependence of the scattered  intensity, permits determination of the matrix elements in terms of a well-known fiducial quantity, which can be the matrix elements of the principal resonance transition.   The approach is appealing because it can lead to high relative precision ($<10^{-3}$) in the matrix elements,  and because it is a frequency domain method which describes spectroscopy of energy level locations and of matrix elements in a uniform manner.

\section{Acknowledgments}

Useful discussions with A. Sieradzan, A. Beger, W. vanOrden, and G. Hoy are acknowledged.  The financial support of the National Science Foundation (Grant NSF-PHY-9504864) is greatly appreciated.


\begin{thebibliography}{99}

\bibitem{Tannoudji} Claude Cohen-Tannoudji, Jacques Dupont-Rec, and Gilbert Grynberg, Atom-Photon                      Interactions: Basic Processes and Applications, (John Wiley and Sons, Inc., New York, 1992); Rodney Louden, Quantum Theory of Light, 2nd Ed. (Oxford University Press, 1983).

\bibitem{Noecker} M.C. Noecker, B.P. Masterson, and C.E. Wieman, Phys. Rev. Lett. 61, 310 (1988); M.A. Bouchiat, Atomic Physics 12, AIP Conference Proceedings 232, ed. J.C. Zorn and R.R. Lewis,  AIP (New York, 1990).

\bibitem{Mihalas}D. Mihalas, Stellar Atmospheres, (Freeman, San Francisco, 1970).

\bibitem{Goebel} Dirk Goebel and Uwe Hohm, Phys. Rev A52, 3691 (1995);  M.A.Kadar-Kallen and K.D. Bonin, Phys. Rev. Lett. 72, 828 (1994).	

\bibitem{Ekstrom} C.R. Ekstrom, J. Schmiedmayer, M.S. Chapman, T.D. Hammond, and D.E. Pritchard, Phys. Rev. A51, 3883 (1995); W.I. McAlexander, E.R.I. Abraham, N.W.M. Ritchie, C.J. Williams, H.T.C. Stoof,  and R.G. Hulet, Phys. Rev. A51, R871 (1995); H. Wang, J. Li, X.T. Wang, C.J. Williams, P.L. Gould, and W.C. Stwalley, Phys. Rev. A55, R1569 (1997).

\bibitem{Oates} C.W. Oates, K.R. Vogel, and J.L. Hall, Phys. Rev. Lett. 76, 2866 (1996).

\bibitem{Rafec} R.J. Rafec, C.E. Tanner, A.E. Livingston, K.W. Kulka, J.G. Berry, and C.A. Kurtz, Phys. Rev. A 50, R1976 (1994); L. Young,  W.T. Hill III, S.J. Sibener, S.D. Price, C.E. Tanner, C.E.  Wieman, and S.R. Leone, Phys. Rev. A50 2174 (1994).

\bibitem{Volz} U. Volz, M.Jajerus, H. Liebel, A. Schmitt, and H. Schmoranzer, Phys. Rev. Lett. 76, 2862
      (1996).  U. Volz and H. Schmoranzer, Physica Scripta T65, 48 (1996).

\bibitem{Meyer} R.P. Meyer, A.I. Beger, and M.D. Havey, Phys. Rev. A55, 230  (1997).

\bibitem{Shao} P. Shao, W. Lichten, H. Layes, J. Bergquist, Phys. Rev. Lett.  58, 1293 (1987); T. W.
      Hänsch, "Precision Spectroscopy of Atomic Hydrogen," in Atomic Physics 14, ed.
      Wineland, C.E. Wieman, and S.J. Smith, AIP Conference Proceedings  (New York, 1995).

\bibitem{DiFillipo} F. DiFillipo, V. Natrajan, K.R. Boyce, and D.E. Pritchard,  Phys. Rev. Lett. 73 1481                       (1994).

\bibitem{Marcuse} Dietrich Marcuse, Principles of Quantum Electronics, (Academic Press, New York, 1980).

\bibitem{Quattropani} A. Quattropani, F. Bassani and Sandra Carillo, Phys. Rev. A25, 3079 (1982); J.H. Tung,                A.Z. Tang, G.J. Salamo, and F.T. Chan, J. Opt. Soc. Am. B3, 837 (1986); J.H. Tung, X.M. Ye, G.J. Salamo, and F.T. Chan, Phys. Rev. A30, 1175 (1984); C.M. Penney, J. Opt. Soc. Am. 59, 34 (1969); J.E. Bjorkholm and P.F. Liao, Phys. Rev. Lett. 33, 128 (1974).

\bibitem{Tam} A.C. Tam and C.K. Au, Optics Comm. 19, 265 (1976).

\bibitem{Walkup} R. Walkup, A.L. Migdall, and D.E. Pritchard, Phys. Rev. A25, 3114 (1982).

\bibitem{Zei} D. Zei, R.N. Compton, J. Stockdale, M. Pindzola, Phys. Rev. A40, 5044 (1989).

\bibitem{Beger} A. I. Beger, M.D. Havey, and R.P. Meyer, Phys. Rev. A 55, 3780 (1997).

\bibitem{Bethe} H.A Bethe and E.E. Salpeter, Quantum Mechanics of One- and Two-Electron Atoms,
       (Springer-Verlag, Berlin, 1957).

\bibitem{Weise} W.L. Weise, M.W. Smith, B.M. Miles, Atomic Transition Probabilities: Sodium
      Through Calcium, Natl. Bur. Stand. (U.S. GPO, Washington, D.C. 1966, 1969), V. 1-2;
      A. Lindgård and S.E. Hielson, Atomic Data and Nuclear Data Tables, v.19 (Academic
      Press, 1977);  D.R. Bates and A. Damgaard, Phil. Tran. Roy, Soc. Lon. 242, 101 (1949).

\bibitem{Theo} C.E. Theodosiou, I.J. Curtis, C.A. Nicolaides, Phys. Rev. A 52, 3677 (1995).

\bibitem{Beck} D.R. Beck, C.A. Nicolaides, Chem. Phys. Lett. 49, 357 (1977).

\bibitem{Radzig} A.A. Radzig and B.M. Smirnov, Reference Data on Atoms, Molecules, and Ions (Springer-
       Verlag, Berlin, 1985).

\bibitem{Definition1} The oscillator strength is related to the squared matrix elements by $f_{ik} = 2m\omega_{ik}|<i|qr|k>|^2 /3\hbar (2J_i + 1)$, where q and m are the electron mass and charge, and $J_i$ the angular momentum of the level i.

\bibitem{Definition2} For a ground multiplet split by the spin-orbit interaction, detection of Raman scattered
       light, rather than the Rayleigh scattered radiation, can greatly improve the signal-to-background ratio.

\end{thebibliography}
\end{document}